\documentclass[12pt]{article}

\usepackage{epsfig,amsmath,amssymb,amsfonts,amstext,amsthm}
\usepackage{latexsym,graphics,epsf,epsfig,color}
\usepackage{url}

\newtheorem{theorem}{Theorem}

\newtheorem{coro}[theorem]{Corollary}

\begin{document}

\title{Security Embedding Codes}

\author{Hung D. Ly, Tie Liu, and Yufei Blankenship\thanks{This research was supported in part by the National Science Foundation under Grant CCF-09-16867 and by a gift grant from the Huawei Technologies USA. The material of this paper was presented in part at the 2010 IEEE International Symposium on Information Theory, Austin, TX, June 2010. Hung D. Ly and Tie Liu are with the Department of Electrical and Computer Engineering, Texas A\&M University, College Station, TX 77843, USA (e-mail: \{hungly,tieliu\}@tamu.edu). Yufei Blankenship is with the Huawei Technologies, Rolling Meadows, IL 60008, USA (e-mail: yblankenship@huawei.com).}
}

\date{\today}

\maketitle

\begin{abstract}
This paper considers the problem of simultaneously communicating two messages, a high-security message and a low-security message, to a legitimate receiver, referred to as the security embedding problem. An information-theoretic formulation of the problem is presented. A coding scheme that combines rate splitting, superposition coding, nested binning and channel prefixing is considered and is shown to achieve the secrecy capacity region of the channel in several scenarios. Specifying these results to both scalar and independent parallel Gaussian channels (under an average individual per-subchannel power constraint), it is shown that the high-security message can be embedded into the low-security message at full rate (as if the low-security message does not exist) without incurring any loss on the overall rate of communication (as if both messages are low-security messages). Extensions to the wiretap channel II setting of Ozarow and Wyner are also considered, where it is shown that ``perfect" security embedding can be achieved by an encoder that uses a two-level coset code.
\end{abstract}

\section{Introduction}
Physical layer security has been a very active area of research in information theory. See \cite{LPS-M09} and \cite{LT-M10} for overviews of recent progress in this field. A basic model of physical layer security is a wiretap/broadcast channel  \cite{Wyn-BSTJ75,CK-IT78} with two receivers, a legitimate receiver and an eavesdropper. Both the legitimate receiver and the eavesdropper channels are assumed to be \emph{known} at the transmitter. By exploring the (statistical) difference between the legitimate receiver channel and the eavesdropper channel, one may design coding schemes that can deliver a message reliably to the legitimate receiver while keeping it asymptotically perfectly secret from the eavesdropper.

While assuming the transmitter's knowledge of the legitimate receiver channel might be reasonable (particularly when a feedback link is available), assuming that the transmitter knows the eavesdropper channel is \emph{unrealistic} in most scenarios. This is mainly because the eavesdropper is an \emph{adversary}, who usually has no incentive to help the transmitter to acquire its channel state information. Hence, it is critical that physical layer security techniques are designed to withstand the \emph{uncertainty} of the eavesdropper channel.

In this paper, we consider a communication scenario where there are \emph{multiple} possible realizations for the eavesdropper channel. Which realization will actually occur is \emph{unknown} to the transmitter. Our goal is to design coding schemes such that the number of \emph{secure} bits delivered to the legitimate receiver depends on the \emph{actual} realization of the eavesdropper channel. More specifically, when the eavesdropper channel realization is weak, \emph{all} bits delivered to the legitimate receiver need to be secure. In addition, when the eavesdropper channel realization is strong, a prescribed \emph{part} of the bits needs to \emph{remain} secure. We call such codes \emph{security embedding codes}, referring to the fact that high-security bits are now embedded into the low-security ones. We envision that such codes are naturally useful for the secrecy communication scenarios where the information bits are \emph{not} created equal: some of them have more security priorities than the others and hence require stronger security protections during communication. For example, in real wireless communication systems, control plane signals have higher secrecy requirement than data plane transmissions, and signals that carry users' identities and cryptographic keys require stronger security protections than the other signals.

A key question that we consider is at what expense one may allow part of the bits to enjoy stronger security protections. Note that a ``naive" security embedding scheme is to design two separate secrecy codes to provide two different levels of security protections, and apply them to two separate parts of the information bits. In this scheme, the high-security bits are protected using a stronger secrecy code and hence are communicated at a lower rate. The overall communication rate is a \emph{convex} combination of the low-security bit rate and the high-security bit rate and hence is lower than the low-security bit rate. Moreover, this rate loss becomes larger as the portion of the high-security bits becomes larger and the additional security requirement (for the high-security bits) becomes higher. 

The main result of this paper is to show that it is possible to have a significant portion of the information bits enjoying additional security protections \emph{without} sacrificing the overall communication rate. This further justifies the name ``security embedding," as having part of the information bits enjoying additional security protections is now only an added bonus. More specifically, in this paper, we call a secrecy communication scenario \emph{embeddable} if a \emph{nonzero} fraction of the information bits can enjoy additional security protections without sacrificing the overall communication rate, and we call it \emph{perfectly embeddable} if the high-security bits can be communicated at \emph{full} rate (as if the low-security bits do not exist) without sacrificing the overall communication rate. Key to achieve optimal security embedding is to \emph{jointly} encode the low-security and high-security bits (as opposed to separate encoding as in the naive scheme). In particular, the low-security bits can be used as (part of) the \emph{transmitter randomness} to protect the high-security bits (when the eavesdropper channel realization is strong); this is a key feature of our proposed security embedding codes.

The rest of the paper is organized as follows. In Sec.~\ref{sec:wtc}, we briefly review some basic results on the secrecy capacity and optimal encoding scheme for several classical wiretap channel settings. These results provide performance and structural benchmarks for the proposed security embedding codes. In Sec.~\ref{sec:mswtc}, an information-theoretic formulation of the security embedding problem is presented, which we term as \emph{two-level security wiretap channel}. A coding scheme that combines rate splitting, superposition coding, nested binning and channel prefixing is proposed and is shown to achieve the secrecy capacity region of the channel in several scenarios. Based on the results of Sec.~\ref{sec:mswtc}, in Sec.~\ref{sec:gmswtc} we study the engineering communication models with real channel input and additive white Gaussian noise, and show that both scalar and independent parallel Gaussian (under an individual per-subchannel average power constraint) two-level security wiretap channels are \emph{perfectly embeddable}. In Sec.~\ref{sec:mswtc2}, we extend the results of Sec.~\ref{sec:mswtc} to the \emph{wiretap channel II} setting of Ozarow and Wyner \cite{OW-BSTJ84}, and show that two-level security wiretap channels II are also \emph{pefectly embeddable}. Finally, in Sec.~\ref{sec:con}, we conclude the paper with some remarks.  

\section{Wiretap Channel: A Review} \label{sec:wtc}

\begin{figure}[t]
\begin{center}
\scalebox{0.5}{\includegraphics{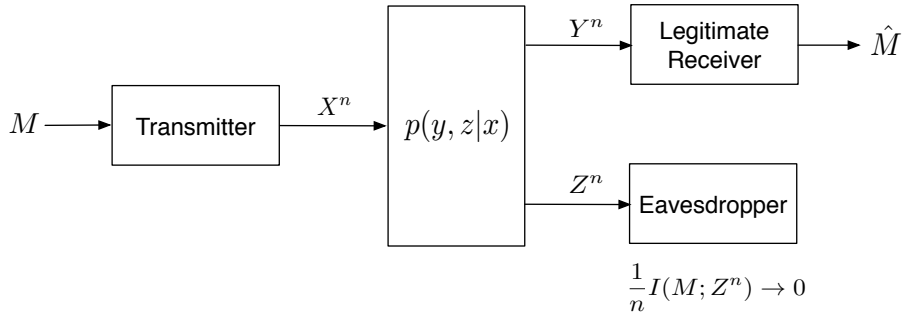}}
\caption{Wiretap channel.}
\label{fig:wtc}
\end{center}
\end{figure}

Consider a discrete memoryless wiretap channel with transition probability $p(y,z|x)$, where $X$ is the channel input, and $Y$ and $Z$ are the channel outputs at the legitimate receiver and the eavesdropper, respectively (see Fig.~\ref{fig:wtc}). The transmitter has a message $M$, uniformly drawn from $\{1,\ldots,2^{nR}\}$ where $n$ is the block length and $R$ is the rate of communication. The message $M$ is intended for the legitimate receiver, but needs to be kept asymptotically perfectly secret from the eavesdropper. Mathematically, this secrecy constraint can be written as
\begin{equation}
\frac{1}{n}I(M;Z^n) \rightarrow 0
\label{eq:cons1}
\end{equation}
in the limit as $n \rightarrow \infty$, where $Z^n=(Z[1],\ldots,Z[n])$ is the collection of the channel outputs at the eavesdropper during communication. A communication rate $R$ is said to be \emph{achievable} if there exists a sequence of codes of rate $R$ such that the message $M$ can be reliably delivered to the legitimate receiver while satisfying the asymptotic perfect secrecy constraint \eqref{eq:cons1}. The largest achievable rate is termed as the \emph{secrecy capacity} of the channel.

A discrete memoryless wiretap channel $p(y,z|x)$ is said to be \emph{degraded} if $X \rightarrow Y \rightarrow Z$ forms a Markov chain in that order. The secrecy capacity $C_s$ of a degraded wiretap channel was characterized by Wyner \cite{Wyn-BSTJ75} and can be written as
\begin{equation}
C_s = \max_{p(x)} \left[I(X;Y)-I(X;Z)\right]
\label{eq:Cs-Wyn}
\end{equation}
where the maximization is over all possible input distributions $p(x)$. The scheme proposed in \cite{Wyn-BSTJ75} to achieve the secrecy capacity \eqref{eq:Cs-Wyn} is \emph{random binning}, which can be described as follows. 

Consider a codebook of $2^{n(R+T)}$ codewords, each of length $n$. The codewords are partitioned into $2^{nR}$ bins, each containing $2^{nT}$ codewords. Given a message $m$ (which is uniformly drawn from $\{1,\ldots,2^{nR}\}$), the encoder \emph{randomly} and uniformly chooses a codeword $x^n$ in the $m$th bin and sends it through the channel. The legitimate receiver needs to decode the entire codebook (and hence recover the transmitted message $m$), so the overall rate $R+T$ cannot be too high. On the other hand, the rate $T$ of the sub-codebooks in each bin represents the amount of external randomness injected by the transmitter (transmitter randomness) into the channel and hence needs to be sufficiently large to confuse the eavesdropper. With an appropriate choice of the codebooks and the partitions of bins, it was shown in \cite{Wyn-BSTJ75} that any communication rate $R$ less than the secrecy capacity \eqref{eq:Cs-Wyn} is achievable by the aforementioned random binning scheme.

For a \emph{general} discrete memoryless wiretap channel $p(y,z|x)$ where the channel outputs $Y$ and $Z$ are \emph{not} necessarily ordered, the random binning scheme of \cite{Wyn-BSTJ75} is \emph{not} necessarily optimal. In this case, the secrecy capacity $C_s$ of the channel was characterized by Csisz\'{a}r and K\"{o}rner \cite{CK-IT78} and can be written as
\begin{equation}
C_s = \max_{p(v,x)} \left[I(V;Y)-I(V;Z)\right]
\label{eq:Cs-CK}
\end{equation}
where $V$ is an auxiliary random variable satisfying the Markov chain $V \rightarrow X \rightarrow (Y,Z)$. The scheme proposed in \cite{CK-IT78} is to first \emph{prefix} the channel input $X$ by $V$ and view $V$ as the input of the \emph{induced} wiretap channel $p(y,z|v)=\sum_{x}p(y,z|x)p(x|v)$. Applying the random binning scheme of \cite{Wyn-BSTJ75} to the induced wiretap channel $p(y,z|v)$ proves the achievability of rate $I(V;Y)-I(V;Z)$ for any given joint auxiliary-input distribution $p(v,x)$.

In communication engineering, communication channels are usually modeled as discrete-time channels with real input and additive white Gaussian noise. Consider a (scalar) Gaussian wiretap channel where the channel outputs at the legitimate receiver and the eavesdropper are given by
\begin{equation}
\begin{array}{rcl}
Y & = & \sqrt{a}X+N_1\\
Z & = & \sqrt{b}X+N_2. 
\end{array}
\label{eq:Ch-GWTC}
\end{equation}  
Here, $X$ is the channel input which is subject to the average power constraint 
\begin{equation}
\frac{1}{n}\sum_{i=1}^n (X[i])^2 \leq P
\label{eq:APC}
\end{equation}
$a$ and $b$ are the channel gains for the legitimate receiver and the eavesdropper channel respectively, and $N_1$ and $N_2$ are additive white Gaussian noise with zero means and \emph{unit} variances. The secrecy capacity of the channel was characterized in \cite{LH-IT78} and can be written as
 \begin{equation}
C_s(P,a,b) = \left[\frac{1}{2}\log(1+aP)-\frac{1}{2}\log(1+bP)\right]^+
\label{eq:Cs-LH}
\end{equation}
where $[x]^+:=\max(0,x)$. Note from \eqref{eq:Cs-LH} that $C_s(P,a,b)>0$ if and only if $a>b$. That is, for the Gaussian wiretap channel \eqref{eq:Ch-GWTC}, asymptotic perfect secrecy communication is possible if and only if the legitimate receiver has a larger channel gain than the eavesdropper. In this case, we can equivalently write the channel output $Z$ at the eavesdropper as a degraded version of the channel output $Y$ at the legitimate receiver, and the random binning scheme of \cite{Wyn-BSTJ75} with \emph{Gaussian} codebooks and \emph{full} transmit power achieves the secrecy capacity of the channel.

A closely related engineering scenario consists of a bank of $L$ independent parallel scalar Gaussian wiretap channels \cite{LYT-All06}. In this scenario, the channel outputs at the legitimate receiver and the eavesdropper are given by $Y=(Y_1,\ldots,Y_L)$ and $Z=(Z_1,\ldots,Z_L)$ where
\begin{equation}
\begin{array}{rcl}
Y_l & = & \sqrt{a_l}X_l+N_{1,l}\\
Z_l & = & \sqrt{b_l}X_l+N_{2,l} 
\end{array}, \quad l=1,\ldots,L.
\label{eq:Ch-PGWTC}
\end{equation}  
Here, $X_l$ is the channel input for the $l$th subchannel, $a_l$ and $b_l$ are the channel gains for the legitimate receiver and the eavesdropper channel respectively in the $l$th subchannel, and $N_{1,l}$ and $N_{2,l}$ are additive white Gaussian noise with zero means and \emph{unit} variances. Furthermore, $(N_{1,l},N_{2,l})$ are independent for $l=1,\ldots,L$ so all $L$ subchannels are independent of each other. 

Two different types of power constraints have been considered: the average individual per-subchannel power constraint
\begin{equation}
\frac{1}{n}\sum_{i=1}^n (X_l[i])^2 \leq P_l, \quad l=1,\ldots,L
\label{eq:pcons-s}
\end{equation}
and the average total power constraint
\begin{equation}
\sum_{l=1}^{L}\left[\frac{1}{n}\sum_{i=1}^n (X_l[i])^2\right] \leq P.
\label{eq:pcons-t}
\end{equation}
Under the average individual per-subchannel power constraint \eqref{eq:pcons-s}, the secrecy capacity of the independent parallel Gaussian wiretap channel \eqref{eq:Ch-PGWTC} is given by \cite{LYT-All06}
\begin{equation}
C_s(\{P_l,a_l,b_l\}_{l=1}^{L})=\sum_{l=1}^{L}C_s(P_l,a_l,b_l)
\label{eq:Cs-LYT}
\end{equation}
where $C_s(P,a,b)$ is defined as in \eqref{eq:Cs-LH}. Clearly, any communication rate less than the secrecy capacity \eqref{eq:Cs-LYT} can be achieved by using $L$ separate scalar Gaussian wiretap codes, each for one of the $L$ subchannels. The secrecy capacity, $C_s(P,\{a_l,b_l\}_{l=1}^{L})$, under the average total power constraint \eqref{eq:pcons-t} is given by
\begin{equation}
C_s(P,\{a_l,b_l\}_{l=1}^{L}) 
=\max_{(P_1,\ldots,P_L)}\sum_{l=1}^{L}C_s(P_l,a_l,b_l)
\end{equation}
where the maximization is over all possible power allocations $(P_1,\ldots,P_L)$ such that $\sum_{l=1}^{L}P_l \leq P$. A waterfilling-like solution for the optimal power allocation was derived in \cite[Th.~1]{LYT-All06}, which provides an efficient way to numerically calculate the secrecy capacity $C_s(P,\{a_l,b_l\}_{l=1}^{L})$.

\section{Two-Level Security Wiretap Channel} \label{sec:mswtc}
\subsection{Channel Model}\label{sec:mswtc-mod}
Consider a discrete memoryless broadcast channel with three receivers and transition probability $p(y,z_1,z_2|x)$. The receiver that receives the channel output $Y$ is a legitimate receiver. The receivers that receive the channel outputs $Z_1$ and $Z_2$ are two possible realizations of an eavesdropper. Assume that the channel output $Z_2$ is \emph{degraded} with respect to the channel output $Z_1$, i.e.,
\begin{equation}
X \rightarrow Z_1 \rightarrow Z_2 \label{eq:mk}
\end{equation}
forms a Markov chain in that order. Therefore, the receiver that receives the channel output $Z_1$ represents a stronger realization of the eavesdropper channel than the receiver that receives the channel output $Z_2$. 

\begin{figure}[t]
\begin{center}
\scalebox{0.5}{\includegraphics{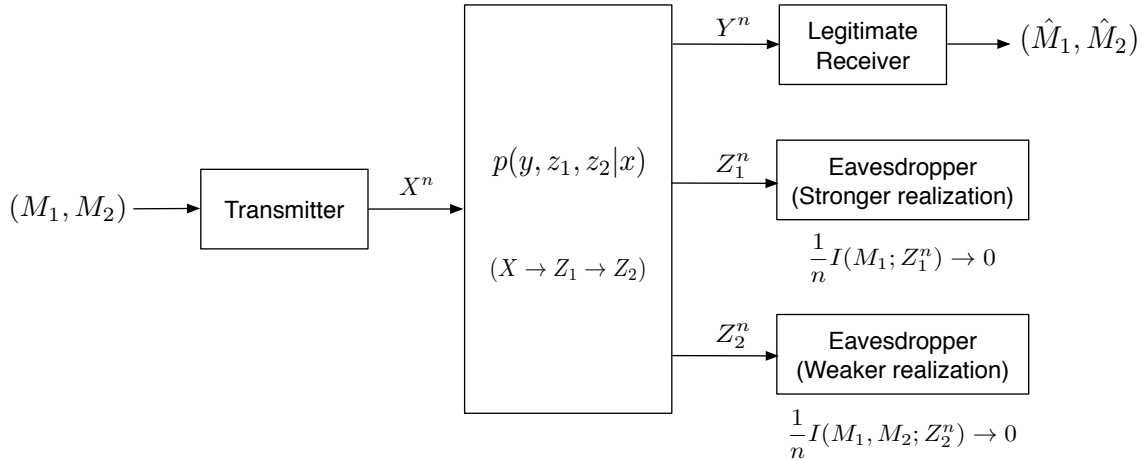}}
\caption{Two-level security wiretap channel.}
\label{fig:mswtc}
\end{center}
\end{figure}

The transmitter has two independent messages: a high-security message $M_1$ uniformly drawn from $\{1,\ldots,2^{nR_1}\}$ and a low-security message $M_2$ uniformly drawn from $\{1,\ldots,2^{nR_2}\}$. Here, $n$ is the block length, and $R_1$ and $R_2$ are the corresponding rates of communication. Both messages $M_1$ and $M_2$ are intended for the legitimate receiver, and need to be kept asymptotically perfectly secure when the eavesdropper realization is weak, i.e.,
\begin{equation}
\frac{1}{n}I(M_1,M_2;Z_2^n) \rightarrow 0
\label{eq:cons2-1}
\end{equation}
in the limit as $n \rightarrow \infty$. In addition, when the eavesdropper realization is strong, the high-security message $M_1$ needs to remain asymptotically perfectly secure, i.e., 
\begin{equation}
\frac{1}{n}I(M_1;Z_1^n) \rightarrow 0 
\label{eq:cons2-2}
\end{equation}
in the limit as $n \rightarrow \infty$. A rate pair $(R_1,R_2)$ is said to be \emph{achievable} if there is a sequence of codes of rate pair $(R_1,R_2)$ such that both messages $M_1$ and $M_2$ can be reliably delivered to the legitimate receiver while satisfying the asymptotic perfect secrecy constraints \eqref{eq:cons2-1} and \eqref{eq:cons2-2}. The collection of all possible achievable rate pairs is termed as the \emph{secrecy capacity region} of the channel. Fig.~\ref{fig:mswtc} illustrates this communication scenario, which we term as \emph{two-level security wiretap channel}.

The above setting of two-level security wiretap channel is closely related to the traditional wiretap channel setting of \cite{Wyn-BSTJ75,CK-IT78}. More specifically, without the additional secrecy constraint \eqref{eq:cons2-2} on the high-security message $M_1$, we can simply view the messages $M_1$ and $M_2$ as a single (low-security) message $M$ with rate $R_1+R_2$. And the problem reduces to communicating the message $M$ over the traditional wiretap channel with transition probability $p(y,z_2|x)=\sum_{z_1}p(y,z_1,z_2|x)$. By the secrecy capacity expression \eqref{eq:Cs-CK}, the maximum achievable $R_1+R_2$ is given by
\begin{equation}
\max_{p(v,x)}\left[I(V;Y)-I(V;Z_2)\right]
\label{eq:Cs_CK2}
\end{equation}
where $V$ is an auxiliary random variable satisfying the Markov chain $V \rightarrow X \rightarrow (Y,Z_2)$.
Similarly, without needing to communicate the low-security message $M_2$ (i.e., $R_2=0$), the secrecy constraint \eqref{eq:cons2-2} reduces to $(1/n)I(M_1;Z_2^n) \rightarrow 0$ which is implied by the secrecy constraint \eqref{eq:cons2-1} since $I(M_1;Z_2^n) \leq I(M_1;Z_1^n)$ due to the Markov chain \eqref{eq:mk}. In this case, the problem reduces to communicating the high-security message $M_1$ over the traditional wiretap channel with transition probability 
$p(y,z_1|x)=\sum_{z_2}p(y,z_1,z_2|x)$. Again, by the secrecy capacity expression \eqref{eq:Cs-CK}, the maximum achievable $R_1$ is given by
\begin{equation}
\max_{p(w,x)}\left[I(W;Y)-I(W;Z_1)\right]
\label{eq:Cs_CK3}
\end{equation}
where $W$ is an auxiliary random variable satisfying the Markov chain $W \rightarrow X \rightarrow (Y,Z_1)$. 

Based on the above connections, we may conclude that a two-level security wiretap channel $p(y,z_1,z_2|x)$ is \emph{embeddable} if there exists a sequence of coding schemes with a rate pair $(R_1,R_2)$ such that $R_1+R_2$ is equal to \eqref{eq:Cs_CK2} and $R_1>0$, and it is \emph{perfectly embeddable} if there exists a sequence of coding schemes with a rate pair $(R_1,R_2)$ such that $R_1+R_2$ is equal to \eqref{eq:Cs_CK2} and $R_1$ is equal to \eqref{eq:Cs_CK3}.

An important special case of the two-level security wiretap channel problem considered here is when the channel output $Z_2$ is a constant signal. In this case, the secrecy constraint \eqref{eq:cons2-1} becomes \emph{obsolete}, and the low-security message $M_2$ becomes a \emph{regular} message without any secrecy constraint. The problem of simultaneously communicating a regular message and a confidential message over a discrete memoryless wiretap channel was first considered in \cite{LLPS-ITS11}, where a single-letter characterization of the capacity region was established. For the \emph{general} two-level security wiretap channel problem that we consider here, both high-security message $M_1$ and low-security message $M_2$ are subject to asymptotic perfect secrecy constraints, which makes the problem potentially much more involved. 

\subsection{Main Results}
The following theorem provides two \emph{sufficient} conditions for establishing the achievability of a rate pair for a given discrete memoryless two-level security wiretap channel.

\begin{theorem} \label{thm:DM1}
Consider a discrete memoryless two-level security wiretap channel with transition
probability $p(y,z_1,z_2|x)$ that satisfies the Markov chain
\eqref{eq:mk}. A nonnegative pair $(R_1,R_2)$ is an achievable rate pair of the channel if it satisfies 
\begin{equation}
\begin{array}{rll}
R_1 & \leq & I(X;Y)-I(X;Z_1)\\
R_1+R_2 & \leq & I(X;Y)-I(X;Z_2)
\end{array}
\label{eq:DM1-1}
\end{equation}
for some input distribution $p(x)$. More generally, a nonnegative pair $(R_1,R_2)$ is an achievable rate pair of the channel if it satisfies
\begin{equation}
\begin{array}{rll}
R_1 & \leq & I(V;Y|U)-I(V;Z_1|U)\\
R_1+R_2 & \leq & I(V;Y)-I(V;Z_2)
\end{array}
\label{eq:DM1-2}
\end{equation}
for some joint distribution $p(u,v,x)$, where $U$ and $V$ are auxiliary random variables satisfying the Markov chain $U \rightarrow V \rightarrow X \rightarrow (Y,Z_1,Z_2)$ and such that $I(U;Y) \geq I(U;Z_2)$.
\end{theorem}

Clearly, the sufficient condition \eqref{eq:DM1-1} can be obtained from \eqref{eq:DM1-2} by choosing $V=X$ and $U$ to be a constant. Hence, \eqref{eq:DM1-2} is a more general sufficient condition than \eqref{eq:DM1-1}. The sufficient condition \eqref{eq:DM1-1} can be proved by considering a \emph{nested} binning scheme that uses the low-security message $M_2$ as part of the transmitter randomness to protect the high-security message $M_1$ (when the eavesdropper channel realization is strong). The more general sufficient condition \eqref{eq:DM1-2} can be proved by considering a more complex coding scheme that combines rate splitting, superposition coding, nested binning and channel prefixing. A detailed proof of the theorem is provided in Sec.~\ref{pf:thm1}.

The following corollary provides sufficient conditions for establishing that a two-level security wiretap channel is (perfectly) embeddable. The conditions are given in terms of the existence of a joint auxiliary-input random triple and are immediate consequences of Theorem~\ref{thm:DM1}.

\begin{coro}
A two-level security wiretap channel $p(y,z_1,z_2|x)$ is \emph{embeddable} if there exists a pair of auxiliary random variables $U$ and $V$ satisfying the Markov chain $U \rightarrow V \rightarrow X \rightarrow (Y,Z_1,Z_2)$ and such that  $I(U;Y) \geq I(U;Z_2)$, $p(v,x)$ is an \emph{optimal} solution to the maximization program \eqref{eq:Cs_CK2}, and $I(V;Y|U)-I(V;Z_1|U)>0$, and it is \emph{perfectly embeddable} if there exists a pair of auxiliary random variables $U$ and $V$ satisfying the Markov chain $U \rightarrow V \rightarrow X \rightarrow (Y,Z_1,Z_2)$ and such that $I(U;Y) \geq I(U;Z_2)$, $p(v,x)$ is an \emph{optimal} solution to the maximization program \eqref{eq:Cs_CK2}, and $I(V;Y|U)-I(V;Z_1|U)$ is equal to \eqref{eq:Cs_CK3}.
\end{coro}

If, in addition to the Markov chain \eqref{eq:mk}, we also have the Markov chain
\begin{equation}
X \rightarrow Y \rightarrow Z_2 \label{eq:mk2}
\end{equation}
in that order, the sufficient condition \eqref{eq:DM1-2} is also necessary, leading to a precise characterization of the secrecy capacity region. The results are summarized in the following theorem; a proof of the theorem can be found in Appendix \ref{App:1}.

\begin{theorem} \label{thm:DM2}
Consider a discrete memoryless two-level security wiretap channel with transition
probability $p(y,z_1,z_2|x)$ that satisfies the Markov chains
\eqref{eq:mk} and \eqref{eq:mk2}. The secrecy capacity region of the channel is given by the set of all nonnegative pairs that satisfy \eqref{eq:DM1-2} for some joint distribution $p(u,v,x)$, where $U$ and $V$ are auxiliary random variables satisfying the Markov chain $U \rightarrow V \rightarrow X \rightarrow (Y,Z_1,Z_2)$.
\end{theorem}

If, in addition to the Markov chains \eqref{eq:mk} and \eqref{eq:mk2}, we also have the Markov chain
\begin{equation}
X \rightarrow Y \rightarrow Z_1 \label{eq:mk3}
\end{equation}
in that order, the (weaker) sufficient condition \eqref{eq:DM1-1} also becomes necessary, leading to a simpler  characterization of the secrecy capacity region (which does \emph{not} involve any auxiliary random variables). The results are summarized in the following theorem; a proof of the theorem can be found in Appendix \ref{App:2}.

\begin{theorem} \label{thm:DM3}
Consider a discrete memoryless two-level security wiretap channel with transition
probability $p(y,z_1,z_2|x)$ that satisfies the Markov chains
\eqref{eq:mk}, \eqref{eq:mk2} and \eqref{eq:mk3}. The secrecy capacity region of the channel is given by the set of all nonnegative pairs that satisfy \eqref{eq:DM1-1} for some input distribution $p(x)$.
\end{theorem}

\subsection{Proof of Theorem~\ref{thm:DM1}} \label{pf:thm1}

\begin{figure}[t]
\begin{center}
\scalebox{0.3}{\includegraphics{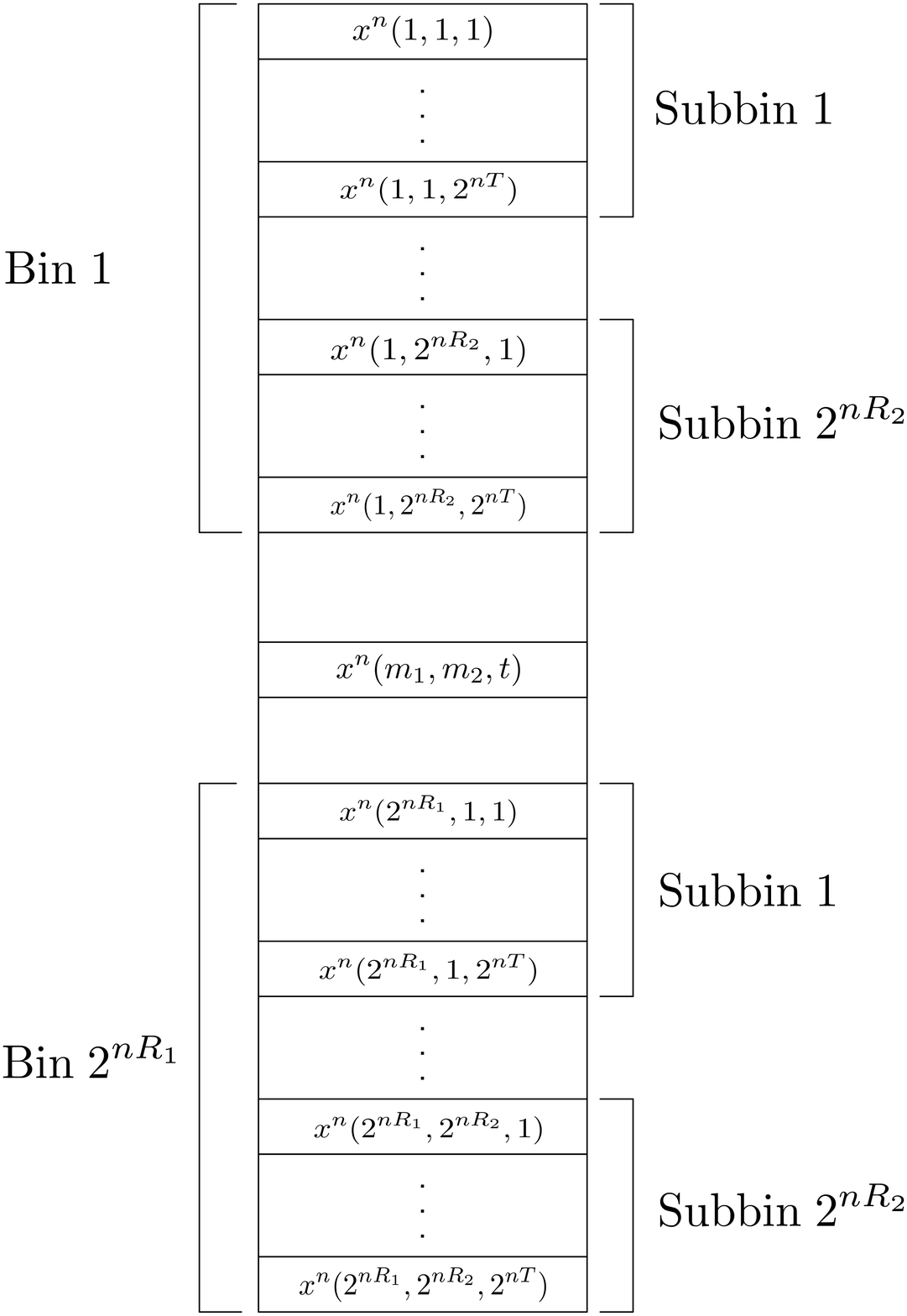}}
\caption{Codebook structure for the nested binning scheme.}
\label{fig:nb}
\end{center}
\end{figure}

We first prove the weaker sufficient condition \eqref{eq:DM1-1} by considering a nested binning scheme that uses the low-security message $M_2$ as part of the transmitter randomness to protect the high-security message $M_1$ (when the eavesdropper channel realization is strong). We shall consider a random-coding argument, which can be described as follows.

Fix an input distribution $p(x)$.

\emph{Codebook generation.} Randomly and
independently generate $2^{n(R_1+R_2+T)}$ codewords of
length $n$ according to an $n$-product of $p(x)$. Randomly partition the codewords
into $2^{nR_1}$ bins so each bin contains
$2^{n(R_2+T)}$ codewords. Further partition each bin into
$2^{nR_2}$ subbins so each subbin contains $2^{nT}$ codewords.
Label the codewords as $x^n_{j,k,l}$ where $j$ denotes
the bin number, $k$ denotes the subbin number within each bin, and
$l$ denotes the codeword number within each subbin. See Fig.~\ref{fig:nb} for an illustration of the codebook structure.

\emph{Encoding.} To send a message pair $(m_1,m_2)$, the transmitter \emph{randomly} (according to a uniform distribution) chooses a codeword $x^n_{m_1,m_2,t}$ from the subbin identified by $(m_1,m_2)$ and sends it through the
channel.

\emph{Decoding at the legitimate receiver.} Given the channel outputs $y^n$, the legitimate receiver looks into
the codebook $\{x^n_{j,k,l}\}_{j,k,l}$ and searches for a codeword that is jointly typical \cite{CT-B91} with $y^n$.
In the case when
\begin{equation}
R_1+R_2+T  < I(X;Y)
\label{eq:FM1}
\end{equation}
with high probability the transmitted codeword $x^n_{m_1,m_2,t}$ is the \emph{only} one that
is jointly typical with $y^n$ (and hence can be correctly decoded).

\emph{Security at the eavesdropper.} Note that each bin corresponds to a message $m_1$ and contains $2^{n(R_2+T)}$ codewords, each randomly and independently generated according to an $n$-product of $p(x)$. For a given message $m_1$, the transmitted codeword is randomly and uniformly chosen from the corresponding bin (where the randomness is from both the low-security message $M_2$ and the transmitter's choice of $t$). Following \cite{Wyn-BSTJ75}, in the case
when
\begin{equation}
R_2+T > I(X;Z_1)
\label{eq:FM2}
\end{equation}
we have $(1/n)I(M_1;Z_1^n)$ tends to zero in the limit as $n \rightarrow 0$. Furthermore, each subbin corresponds to a message pair $(m_1,m_2)$ and contains $2^{nT}$ codewords, each randomly and
independently generated according to an $n$-product of $p(x)$. For a given message pair $(m_1,m_2)$, the transmitted codeword is randomly and uniformly chosen from the corresponding subbin (where the randomness is from the transmitter's choice of $t$). Again, following \cite{Wyn-BSTJ75}, in the case
when
\begin{equation}
T > I(X;Z_2)
\label{eq:FM3}
\end{equation}
we have $(1/n)I(M_1,M_2;Z_2^n)$ tends to zero in the limit as $n \rightarrow 0$.

Eliminating $T$ from \eqref{eq:FM1}--\eqref{eq:FM3} using Fourier-Motzkin elimination, we can conclude that any rate pair $(R_1,R_2)$ that satisfies \eqref{eq:DM1-1} is achievable. 

Next we prove the more general sufficient condition \eqref{eq:DM1-2} by considering a coding scheme that combines rate splitting, superposition coding, nested binning and channel prefixing. We shall once again resort to a random-coding argument, which can be described as follows.

Fix a joint auxiliary-input distribution $p(u)p(v|u)p(x|v)$ with $I(U;Y) \geq I(U;Z_2)$ and $\epsilon>0$. Split the low-security message $M_2$ into two independent submessages $M_2'$ and $M_2''$ with rates $R_2'$ and $R_2''$, respectively.

\begin{figure}[t]
\begin{center}
\scalebox{0.3}{\includegraphics{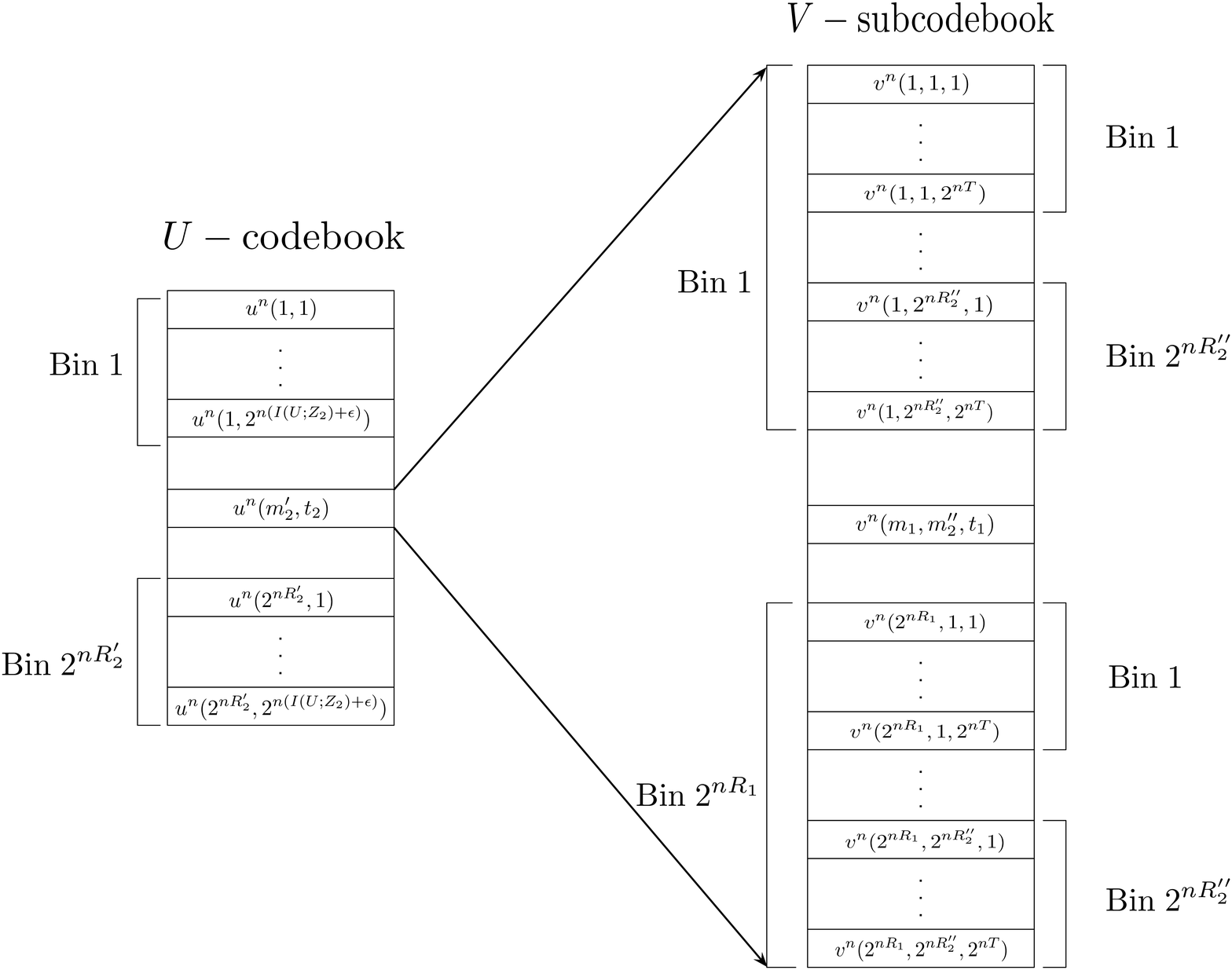}}
\caption{Codebook structure for a coding scheme that combines rate splitting, superposition coding and nested binning.}
\label{fig:nb2}
\end{center}
\end{figure}

\emph{Codebook generation.} Randomly and
independently generate $2^{n(R_2'+I(U;Z_2)+\epsilon)}$ codewords of
length $n$ according to an $n$-product of $p(u)$. Randomly partition the codewords
into $2^{nR_2'}$ bins so each bin contains
$2^{n(I(U;Z_2)+\epsilon)}$ codewords. Label the codewords as
$u^n_{j,k}$ where $j$ denotes the bin number, and $k$ denotes the
codeword number within each bin. We shall refer to the codeword
collection $\{u^n_{j,k}\}_{j,k}$ as the $U$-codebook.

For each codeword $u^n_{j,k}$ in the $U$-codebook, randomly and
independently generate $2^{n(R_1+R_2''+T)}$ codewords of length $n$
according to an $n$-product of $p(v|u)$. Randomly partition
the codewords into $2^{nR_1}$ bins so each bin contains
$2^{n(R_2''+T)}$ codewords. Further partition each bin into
$2^{nR_2''}$ subbins so each subbin contains $2^{nT}$ codewords.
Label the codewords as $v^n_{j,k,l,p,q}$ where $(j,k)$ indicates the base codeword $u_{j,k}^n$ from which $v^n_{j,k,l,p,q}$ was generated, $l$ denotes the bin number, $p$ denotes the subbin number within each bin, and
$q$ denotes the codeword number within each subbin. We shall refer to
the codeword collection $\{v^n_{j,k,l,p,q}\}_{l,p,q}$ as the
$V$-subcodebook corresponding to base codeword $u^n_{j,k}$. See Fig.~\ref{fig:nb2}
for an illustration of the codebook structure.

\emph{Encoding.} To send a message triple $(m_1,m_2',m_2'')$, the
transmitter \emph{randomly} (according a uniform distribution)
chooses a codeword $u^n_{m_2',t_2}$ from the $m_2'$th bin in the
$U$-codebook. Once a $u^n_{m_2',t_2}$ is chosen, the transmitter
looks into the $V$-subcodebook corresponding to $u^n_{m_2',t_2}$ and
\emph{randomly} chooses a codeword $v^n_{m_2',t_2,m_1,m_2'',t_1}$ from the subbin identified by $(m_1,m_2'')$. 
Once a $v^n_{m_2',t_2,m_1,m_2'',t_1}$ is chosen, an input sequence $x^n$ is
generated according to an $n$-product of $p(x|v)$ and is then sent through the
channel.

\emph{Decoding at the legitimate receiver.} Given the channel outputs $y^n$, the legitimate receiver looks into
the $U$-codebook and its $V$-codebooks and searches for a pair of codewords
$(u^n_{j,k},v^n_{j,k,l,p,q})$ that are jointly typical \cite{CT-B91} with $y^n$.
In the case when
\begin{eqnarray}
R_2'+I(U;Z_2)+\epsilon & < & I(U;Y)\label{eq:FM4}\\
\mbox{and} \quad R_1+R_2''+T  & < & I(V;Y|U)\label{eq:FM5}
\end{eqnarray}
with high probability the codeword pair selection
$(u^n_{m_2',t_2},v^n_{m_2',t_2,m_1,m_2'',t_1})$ is the only one that
is jointly typical \cite{CT-B91} with $y^n$.

\emph{Security at the eavesdropper.} To analyze the security of the high-security message $M_1$ and the submessage $M_2''$ at the eavesdropper, we shall assume (for now) that both the submessage $m_2'$ and the codeword selection $u_{m_2',t_2}^n$ are known at the eavesdropper. Note that such an assumption can only \emph{strengthen} our security analysis. Given the base codeword $u_{m_2',t_2}^n$, the encoding of $m_1$ and $m_2''$ using the corresponding $V$-subcodebook is identical to the nested binning scheme considered previously (with additional channel prefixing). Thus in the case when
\begin{eqnarray}
R_2''+T & > & I(V;Z_1|U)\label{eq:FM6}\\
\mbox{and} \quad T & > & I(V;Z_2|U)\label{eq:FM7}
\end{eqnarray}
we have 
\begin{eqnarray}
\frac{1}{n}I(M_1;Z_1^n|M_2') \; = \; \frac{1}{n}I(M_1;Z_1^n,M_2') & \rightarrow & 0 \label{eq:eqv1}\\
\mbox{and} \quad \frac{1}{n}I(M_1,M_2'';Z_2^n|M_2') \; = \;
\frac{1}{n}I(M_1,M_2'';Z_2^n,M_2') & \rightarrow & 0 \label{eq:eqv2}
\end{eqnarray}
in the limit as $n \rightarrow \infty$. The equalities in \eqref{eq:eqv1} and \eqref{eq:eqv2} are due to the fact that $(M_1,M_2'')$ and $M_2'$ are independent. From \eqref{eq:eqv1} we may conclude that $(1/n)I(M_1;Z_1^n) \rightarrow 0$ in the limit as $n \rightarrow \infty$.

To analyze the security of the submessage $M_2'$, note that each bin in the $U$-codebook corresponds to a message $m_2'$ and contains $2^{n(R_2'+I(U;Z_2)+\epsilon)}$ codewords, each randomly and independently generated according to an $n$-product of $p(u)$. For a given submessage $m_2'$, the codeword $u_{m_2',t_2}^n$ is randomly and uniformly chosen from the corresponding bin (where the randomness is from the transmitter's choice of $t_2$). Note from \eqref{eq:FM7} that the rate of each $V$-subcodebook is greater than $I(V;Z_2|U)$. Following \cite[Lemma~1]{CE-ITS09}, we have
\begin{equation}
\frac{1}{n}I(M_2';Z_2^n) \rightarrow 0 \label{eq:eqv3}
\end{equation}
in the limit as $n \rightarrow \infty$.  Putting together \eqref{eq:eqv2} and \eqref{eq:eqv3}, we have
\begin{eqnarray*}
\frac{1}{n}I(M_1,M_2;Z_2^n) &=& \frac{1}{n}I(M_1,M_2',M_2'';Z_2^n)\\
&=&
\frac{1}{n}I(M_2';Z_2^n)+\frac{1}{n}I(M_1,M_2'';Z_2^n|M_2')
\end{eqnarray*}
which tends to zero in the limit as $n \rightarrow \infty$.

Finally, note that the overall communicate rate $R_2$ of the low-security message $M_2$ is given by
\begin{equation}
R_2=R_2'+R_2''. \label{eq:FM8}
\end{equation}
Eliminating $T$, $R_2'$ and $R_2''$ from
\eqref{eq:FM4}--\eqref{eq:FM7}, \eqref{eq:FM8}, and $R_2',R_2'' \geq 0$ using
Fourier-Motzkin elimination, simplifying the results using the
facts that 1) $I(U;Y) \geq I(U;Z_2)$, 2) $I(V;Z_2|U) \leq I(V;Z_1|U)$ which is due to the Markov chain \eqref{eq:mk}, and 3) $I(V;Y|U)+I(U;Y)=I(V,U;Y)=I(V;Y)$  and $I(V;Z_2|U)+I(U;Z_2)=I(V,U;Z_2)=I(V;Z_2)$
which are due to the Markov chain $U \rightarrow V \rightarrow X \rightarrow (Y,Z_1,Z_2)$, and letting $\epsilon \rightarrow 0$, we may conclude that any rate pair $(R_1,R_2)$ satisfying
\eqref{eq:DM1-2} is achievable. This completes the proof of Theorem~\ref{thm:DM1}.

\section{Gaussian Two-Level Security Wiretap Channels} \label{sec:gmswtc}
\subsection{Scalar Channel}
Consider a discrete-time two-level security wiretap channel with real input $X$ and outputs $Y$, $Z_1$ and $Z_2$ given by
\begin{equation}
\begin{array}{rcl}
Y &=& \sqrt{a}X+N_1\\
Z_1 &=& \sqrt{b_1}X+N_{2}\\
Z_2 &=& \sqrt{b_2}X+N_{3}
\end{array}
\label{eq:Ch-MGWTC}
\end{equation}
where $a$, $b_1$ and $b_2$ are the corresponding channel gains, and $N_1$, $N_2$ and $N_3$ are additive white Gaussian noise with zero means and unit variances. Assume that $b_1 \geq b_2$ so the receiver that receives the channel output $Z_1$ represents a stronger realization of the eavesdropper channel than the receiver that receives the channel output $Z_2$. The channel input $X$ is subject to the average power constraint \eqref{eq:APC}.

We term the above communication scenario as \emph{(scalar) Gaussian two-level security wiretap channel}. The following theorem provides an explicit characterization of the secrecy capacity region.

\begin{theorem}\label{thm:MGWTC}
Consider the (scalar) Gaussian two-level security wiretap channel \eqref{eq:Ch-MGWTC}. The secrecy capacity region of the channel is given by the collection of all nonnegative pairs $(R_1,R_2)$ that satisfy
\begin{equation}
\begin{array}{rcl}
R_1 & \leq & C_s(P,a,b_1)\\
\mbox{and} \quad R_1+R_2 & \leq & C_s(P,a,b_2)
\end{array}
\label{eq:Cs-MGWTC}
\end{equation}
where $C_s(P,a,b)$ is defined as in \eqref{eq:Cs-LH}.
\end{theorem}

\emph{Proof:} We first prove the converse part of the theorem. Recall from Sec.~\ref{sec:mswtc-mod} that without transmitting the low-security message $M_2$ (which can only increase the achievable rate $R_1$), the problem reduces to communicating the high-security message $M_1$ over the traditional wiretap channel $p(y,z_1|x)$. For the Gaussian two-level security wiretap channel \eqref{eq:Ch-MGWTC}, the problem reduces to communicating the high-security message $M_1$ over the Gaussian wiretap channel with channel outputs $Y$ and $Z_1$ given by
\begin{equation*}
\begin{array}{rcl}
Y &=& \sqrt{a}X+N_1\\
Z_1 &=& \sqrt{b_1}X+N_{2}.
\end{array}
\end{equation*}
We thus conclude that $R_1 \leq C_s(P,a,b_1)$ for any achievable rate $R_1$. 

Similarly, ignoring the additional secrecy constraint \eqref{eq:cons2-2} for the high-security message $M_1$ (which can only enlarge the achievable rate region $\{(R_1,R_2)\}$), we can simply view the messages $M_1$ and $M_2$ as a single message $M$ with rate $R_1+R_2$. In this case, the problem reduces to communicating the message $M$ over the traditional wiretap channel $p(y,z_2|x)$. For the Gaussian two-level security wiretap channel \eqref{eq:Ch-MGWTC}, the problem reduces to communicating the message $M$ over the Gaussian wiretap channel with channel outputs $Y$ and $Z_2$ given by
\begin{equation*}
\begin{array}{rcl}
Y &=& \sqrt{a}X+N_1\\
Z_2 &=& \sqrt{b_2}X+N_{3}.
\end{array}
\end{equation*}
We thus conclude that $R_1+R_2 \leq C_s(P,a,b_2)$ for any achievable rate pair $(R_1,R_2)$.

To show that any nonnegative pair $(R_1,R_2)$ that satisfies \eqref{eq:Cs-MGWTC} is achievable, let us first consider two simple cases. First, when $b_1 \geq b_2 \geq a$, both $C_s(P,a,b_1)$ and $C_s(P,a,b_2)$ are equal to zero (c.f. definition \eqref{eq:Cs-LH}). So \eqref{eq:Cs-MGWTC} does not include any positive rate pairs and hence there is nothing to prove. Next, when $b_1 \geq a \geq b_2$, $C_s(P,a,b_1)=0$ and \eqref{eq:Cs-MGWTC} reduces to 
\begin{equation*}
\begin{array}{rcl}
R_1& = & 0\\
\mbox{and} \quad R_2 & \leq & C_s(P,a,b_2).
\end{array}
\end{equation*} 
Since the high-security message $M_1$ does not need to be transmitted, any rate pair in this region can be achieved by using a scalar Gaussian wiretap code to encode the low-security message $M_2$. This has left us with the only case with $a \geq b_1 \geq b_2$.

For the case where $a \geq b_1 \geq b_2$, the achievability of any rate pair in \eqref{eq:Cs-MGWTC} follows from that of \eqref{eq:DM1-1} by choosing $X$ to be Gaussian with zero mean and variance $P$. This completes the proof of the theorem. \hfill $\square$

The following corollary follows directly from the achievability of the corner point 
\begin{equation}
(R_1,R_2)=(C_s(P,a,b_1),C_s(P,a,b_2)-C_s(P,a,b_1))
\label{eq:CP}
\end{equation}
of \eqref{eq:Cs-MGWTC}.

\begin{coro}
Scalar Gaussian two-level security wiretap channels under an average power constraint are perfectly embeddable.
\end{coro}

\begin{figure}[t]
\begin{center}
\scalebox{0.5}{\includegraphics{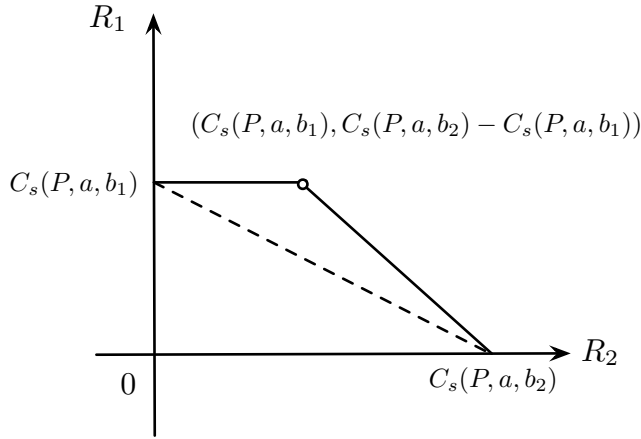}}
\caption{Secrecy capacity region of the scalar Gaussian two-level security wiretap channel ($a>b_1>b_2$). The rate region under the dashed line can be achieved by separate encoding of $M_1$ and $M_2$.}
\label{fig:Cssg}
\end{center}
\end{figure}

Fig.~\ref{fig:Cssg} illustrates the secrecy capacity region \eqref{eq:Cs-MGWTC} for the case where $a > b_1 > b_2$. Also plotted in the figure is the rate region that can be achieved by the naive scheme that uses two Gaussian wiretap codes to encode the messages $M_1$ and $M_2$ separately. Note that the corner point \eqref{eq:CP} is strictly outside the ``naive" rate region, which illustrates the superiority of nested binning over the separate coding scheme.

\subsection{Independent Parallel Channel} 
Consider a discrete-time two-level security wiretap channel which consists of a bank of $L$ independent parallel scalar Gaussian two-level security wiretap channels. In this model, the channel outputs are given by $Y=(Y_1,\ldots,Y_L)$, $Z_1=(Z_{1,1},\ldots,Z_{1,L})$ and $Z_2=(Z_{2,1},\ldots,Z_{2,L})$ where
\begin{equation}
\begin{array}{rcl}
Y_l &=& \sqrt{a_{1}}X_l+N_{1,l}\\
Z_{1,l} &=& \sqrt{b_{1,l}}X_l+N_{2,l}\\
Z_{2,l} &=& \sqrt{b_{2,l}}X_l+N_{3,l}
\end{array} \quad l=1,\ldots,L.
\label{eq:Ch-MPGWTC}
\end{equation}
Here, $X_l$ is the channel input for the $l$th subchannel, $a_l$, $b_{1,l}$ and $b_{2,l}$ are the corresponding channel gains in the $l$th subchannel, and $N_{1,l}$, $N_{2,l}$ and $N_{3,l}$ are additive white Gaussian noise with zero means and unit variances. We assume that $b_{1,l} \geq b_{2,l}$ for all $l=1,\ldots,L$, so the receiver that receives the channel output $Z_1$ represents a stronger realization of the eavesdropper channel in \emph{each} of the $L$ subchannels than the receiver that receives the channel output $Z_2$. Furthermore, $(N_{1,l},N_{2,l},N_{3,l})$, $l=1,\ldots,L$, are independent so all $L$ subchannels are independent of each other. 

We term the above communication scenario as \emph{independent parallel Gaussian two-level security wiretap channel}. The following theorem provides an explicit characterization of the secrecy capacity region under an average individual per-subchannel power constraint.

\begin{theorem} \label{thm:MPGWTC}
Consider the independent parallel Gaussian two-level security wiretap channel \eqref{eq:Ch-MPGWTC} where the channel input $X$ is subject to the average individual per-subchannel power constraint \eqref{eq:pcons-s}. The secrecy capacity region of the channel is given by the collection of all nonnegative pairs $(R_1,R_2)$ that satisfy
\begin{equation}
\begin{array}{rcl}
R_1 & \leq & \sum_{l=1}^{L}C_s(P_l,a_l,b_{1,l})\\
\mbox{and} \quad R_1+R_2 & \leq & \sum_{l=1}^{L}C_s(P_l,a_l,b_{2,l})
\end{array}
\label{eq:Cs-MPGWTC}
\end{equation}
where $C_s(P,a,b)$ is defined as in \eqref{eq:Cs-LH}.
\end{theorem}

\emph{Proof:} We first prove the converse part of the theorem. Following the same argument as that for Theorem~\ref{thm:MGWTC}, we can show that 
\begin{equation}
\begin{array}{rcl}
R_1 & \leq & C_s(\{P_l,a_l,b_{1,l}\}_{l=1}^L)\\
\mbox{and} \quad R_1+R_2 & \leq & C_s(\{P_l,a_l,b_{2,l}\}_{l=1}^L)
\end{array}
\label{T111}
\end{equation}
for any achievable secrecy rate pair $(R_1,R_2)$. By the secrecy capacity expression \eqref{eq:Cs-LYT} for the independent parallel Gaussian wiretap channel under an average individual per-subchannel power constraint, we have 
\begin{equation}
\begin{array}{rcl}
C_s(\{P_l,a_l,b_{1,l}\}_{l=1}^L) &=& \sum_{l=1}^{L}C_s(P_l,a_l,b_{1,l})\\
\mbox{and} \quad C_s(\{P_l,a_l,b_{2,l}\}_{l=1}^L) &=& \sum_{l=1}^{L}C_s(P_l,a_l,b_{2,l}).
\end{array}
\label{T112}
\end{equation}
Substituting \eqref{T112} into \eqref{T111} proves the converse part of the theorem. 

To show that any nonnegative pair $(R_1,R_2)$ that satisfies \eqref{eq:Cs-MPGWTC} is achievable, let us consider \emph{independent} coding over each of the $L$ subchannels. Note that each subchannel is a scalar Gaussian two-level security wiretap channel with average power constraint $P_l$ and channel gains $(a_l,b_{1,l},b_{2,l})$. Thus, by Theorem~\ref{thm:MGWTC}, any nonnegative pair $(R_{1,l},R_{2,l})$ that satisfies 
\begin{equation}
\begin{array}{rcl}
R_{1,l} & \leq & C_s(P_l,a_l,b_{1,l})\\
\mbox{and} \quad R_{1,l}+R_{2,l} & \leq & C_s(P_l,a_l,b_{2,l})
\end{array}
\label{T211}
\end{equation}
is achievable for the $l$th subchannel. The overall communication rates are given by
\begin{equation}
\begin{array}{rcl}
R_1 &=& \sum_{l=1}^{L}R_{1,l}\\
\mbox{and} \quad R_2 &=& \sum_{l=1}^{L}R_{2,l}.
\end{array}
\label{T212}
\end{equation}
Substituting \eqref{T211} into \eqref{T212} proves that any nonnegative pair $(R_1,R_2)$ that satisfies \eqref{eq:Cs-MPGWTC} is achievable. This completes the proof of the theorem. \hfill $\square$  

Similar to the scalar case, the following corollary is an immediate consequence of Theorem~\ref{thm:MPGWTC}.

\begin{coro}
Independent parallel  Gaussian two-level security wiretap channels under an average individual per-subchannel power constraint are perfectly embeddable.
\end{coro}

The secrecy capacity region of the channel under an average total power constraint is summarized in the following corollary. The results follow from the well-known fact that an average total power constraint can be written as the \emph{union} of average individual per-subchannel power constraints, where the union is over all possible power allocations among the subchannels.

\begin{coro}
Consider the independent parallel Gaussian two-level security wiretap channel \eqref{eq:Ch-MPGWTC} where the channel input $X$ is subject to the average total power constraint \eqref{eq:pcons-t}. The secrecy capacity region of the channel is given by the collection of all nonnegative pair $(R_1,R_2)$ that satisfies
\begin{equation}
\begin{array}{rcl}
R_1 & \leq & \sum_{l=1}^{L}C_s(P_l,a_l,b_{1,l})\\
\mbox{and} \quad R_1+R_2 & \leq & \sum_{l=1}^{L}C_s(P_l,a_l,b_{2,l})
\end{array}
\end{equation}
for some power allocation $(P_1,\ldots,P_L)$ such that $\sum_{l=1}^{L}P_l \leq P$.
\end{coro}

\begin{figure}[t]
\begin{center}
\scalebox{0.45}{\includegraphics{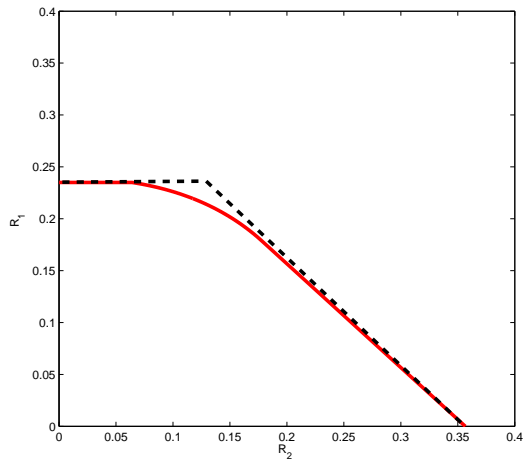}}
\caption{Secrecy capacity region of the independent parallel Gaussian two-level security wiretap channel under an average total power constraint. The intersection of the dashed lines are outside the secrecy capacity region, indicating that the channel is not perfectly embeddable.}
\label{fig:Cspg}
\end{center}
\end{figure}

Fig.~\ref{fig:Cspg} illustrates the secrecy capacity with $L=2$ subchannels where
\begin{eqnarray*}
a_{1} = 1.000, & b_{1,1}=0.800, & b_{2,1}=0.100\\
a_{2} = 1.000, & b_{1,2}=0.250, & b_{2,2}=0.100\\
\mbox{and} \quad P = 1.000.
\end{eqnarray*} 
As we can see, under the average total power constraint \eqref{eq:pcons-t}, the independent parallel Gaussian two-level security wiretap channel is embeddable but \emph{not} perfectly embeddable. The reason is that the optimal power allocation $(P_1,P_2)$ that maximizes $C_s(P_1,a_1,b_{2,1})+C_s(P_2,a_2,b_{2,2})$ is \emph{suboptimal} in maximizing $C_s(P_1,a_1,b_{1,1})+C_s(P_2,a_2,b_{1,2})$. By comparison, under the average individual per-subchannel power constraint \eqref{eq:pcons-s}, the power allocated to each of the subchannels is fixed so the channel is always perfectly embeddable.

\section{Two-Level Security Wiretap Channel II} \label{sec:mswtc2}
In Sec.~\ref{sec:wtc} we briefly summarized the known results on a classical secrecy communication setting known as wiretap channel. A closely related classical secrecy communication scenario is \emph{wiretap channel II}, which was first studied by Ozarow and Wyner \cite{OW-BSTJ84}. In the wiretap channel II setting, the transmitter sends a binary sequence $X^n=(X_1,\ldots,X_n)$ of length $n$ \emph{noiselessly} to an legitimate receiver. The signal $Z^n=(Z_1,\ldots,Z_n)$ received at the eavesdropper is given by 
\begin{equation*}
Z_i = \left\{ 
\begin{array}{ll}
X_i, & i \in S\\
e, & \mbox{otherwise}
\end{array}
\right.
\end{equation*}
where $e$ represents an erasure output, and $S$ is a subset of $\{1,\ldots,n\}$ of size $n\alpha$ representing the locations of the transmitted bits that can be accessed by the eavesdropper.

If the subset $S$ is \emph{known} at the transmitter, a message $M$ of $n(1-\alpha)$ bits can be noiselessly communicated to the legitimate receiver through $X_{S^c}:=\{X_i: \; i \in S^c\}$. Since the eavesdropper has no information regarding to $X_{S^c}$, \emph{perfectly} secure communication is achieved \emph{without} any coding. It is easy to see that  in this scenario, $n(1-\alpha)$ is also the \emph{maximum} number of bits that can be reliably and perfectly securely communicated through $n$ transmitted bits. 

An interesting result of \cite{OW-BSTJ84} is that for any $\epsilon>0$, a total of $n(1-\alpha-\epsilon)$ bits can be reliably and \emph{asymptotically perfectly} securely communicated to the legitimate receiver even when the subset $S$ is \emph{unknown} (but with a fixed size $n\alpha$) a priori at the transmitter. Here, by ``asymptotically perfectly securely" we mean $(1/n)I(M;Z^n) \rightarrow 0$ in the limit as $n \rightarrow \infty$. Unlike the case where the subset $S$ is known , coding is \emph{necessary} when $S$ is unknown a priori at the transmitter. In particular, \cite{OW-BSTJ84} considered a random binning scheme that partitions the collection of all length-$n$ binary sequences into an appropriately chosen \emph{group code} and its cosets. For the wiretap channel setting, as shown in Sec.~\ref{sec:mswtc}, a random binning scheme can be easily modified into a nested binning scheme to efficiently embed high-security bits into low-security ones. The main goal of this section is to extend this result from the classical setting of wiretap channel to wiretap channel II. 


More specifically, assume that a realization of the subset $S$ has two possible sizes, $n\alpha_1$ and $n\alpha_2$, where $1 \geq \alpha_1 \geq \alpha_2 \geq 0$. The transmitter has two independent messages, the high-security message $M_1$ and the low-security message $M_2$, uniformly drawn from $\{1,\ldots,2^{nR_1}\}$ and $\{1,\ldots,2^{nR_2}\}$ respectively. When the size of the realization $S$ is $n\alpha_2$, both messages $M_1$ and $M_2$ need to be secure, i.e., $(1/n)I(M_1,M_2;Z^n) \rightarrow 0$ in the limit as $n \rightarrow \infty$. In addition, when the size of the realization of $S$ is $n\alpha_1$, the high-security message $M_1$ needs to remain secure, i.e., $(1/n)I(M_1;Z^n) \rightarrow 0$ in the limit as $n \rightarrow \infty$. We term this communication scenario as \emph{two-level security wiretap channel II}, in line with our previous terminology in Sec.~\ref{sec:mswtc}. 

By the results of \cite{Wyn-BSTJ75}, without needing to communicate the low-security message $M_2$, the maximum achievable $R_1$ is $1-\alpha_1$. Without the additional secrecy constraint $(1/n)I(M_1;Z^n) \rightarrow 0$ on the high-security message $M_1$, the messages $(M_1,M_2)$ can be viewed as a single message $M$ with rate $R_1+R_2$, and the maximum achievable $R_1+R_2$ is $1-\alpha_2$. The main result of this section is to show that the rate pair $(1-\alpha_1,\alpha_1-\alpha_2)$ is indeed achievable, from which we may conclude that two-level security wiretap channels II are \emph{perfectly} embeddable. Moreover, perfect embedding can be achieved by a nested binning scheme that uses a \emph{two-level} coset code. The results are summarized in the following theorem.

\begin{theorem}
Two-level security wiretap channels II are perfectly embeddable. Moreover, perfect embedding can be achieved by a nested binning scheme that uses a two-level coset code.
\end{theorem}

\emph{Proof:} Fix $\epsilon>0$. Consider a binary parity-check matrix 
$$H=
\left[
\begin{array}{c}
  H_1   \\
  H_2   
\end{array}
\right]
$$
where the size of $H_1$ is $n(1-\alpha_1-\epsilon)\times n$ and the size of $H_2$ is $n(\alpha_1-\alpha_2)\times n$. Let $s_1(\cdot)$ be a one-on-one mapping between $\{1,\ldots,2^{n(1-\alpha_1-\epsilon)}\}$ and the binary vectors of length $n(1-\alpha_1-\epsilon)$, and let $s_2(\cdot)$ be a one-on-one mapping between $M_2\in \{1,\ldots,2^{n(\alpha_1-\alpha_2)}\}$ and the binary vectors of length $n(\alpha_1-\alpha_2)$.

For a given message pair $(m_1,m_2)$, the transmitter randomly (according to a uniform distribution) chooses a solution $x^n$ to the linear equations 
\begin{equation}
(x^n)^tH=(x^n)^t\left[
\begin{array}{c}
  H_1   \\
  H_2   
\end{array}
\right]=
\left[
\begin{array}{c}
  s_1(m_1)   \\
  s_2(m_2)
\end{array}
\right]
\label{eq:gc}
\end{equation}
and sends it to the legitimate receiver. 

When the parity-check matrix $H$ has \emph{full} (row) rank, the above encoding procedure is equivalent of a nested binning scheme that partitions the collection of all length-$n$ binary sequences into bins and subbins using a two-level coset code with parity-check matrices $(H_1,H_2)$. Moreover, let $b_1,\ldots,b_n$ be the columns of $H$ and let $\Gamma \subseteq \{1,\ldots,n\}$. Define $D_2(\Gamma)$ as the dimension of the subspace spanned by $\{b_i: \; i \in \Gamma\}$ and 
\begin{equation*}
D_2^* := \min_{|\Gamma|=n(1-\alpha_2)} D_2(\Gamma).
\end{equation*}
When the size of the realization of $S$ is $n\alpha_2$, by \cite[Lemma~4]{OW-BSTJ84} we have
\begin{equation}
H(M_1,M_2|Z^n)=D_2^*.
\label{BB}
\end{equation}
Note that the low-security message $M_2$ is uniformly drawn from $\{1,\ldots,2^{n(\alpha_1-\alpha_2)}\}$. So by \eqref{eq:gc}, for a given high-security message $m_1$, the transmitted sequence $x^n$ is randomly chosen (according to a uniform distribution) as a solution to the linear equations $(x^n)^tH_1=s_1(m_1)$. If we let $a_1,\ldots,a_n$ be the columns of $H_1$ and define
\begin{equation*}
D_1^* := \min_{|\Gamma|=n(1-\alpha_1)} D_1(\Gamma)
\end{equation*}
where $D_1(\Gamma)$ is the dimension of the subspace spanned by $\{a_i: \; i \in \Gamma\}$, we have again from \cite[Lemma~4]{OW-BSTJ84}
\begin{equation}
H(M_1|Z^n)=D_1^*
\label{AA}
\end{equation}
when the size of the realization of $S$ is $n\alpha_1$.

Let $\Psi(H)=1$ when we have either $H$ does \emph{not} have full rank, or $D_2^* < n(1-\alpha_2-\epsilon)-3/\epsilon$, or $D_1^* < n(1-\alpha_1-\epsilon)-3/\epsilon$, and let $\Psi(H)=0$ otherwise. By using a randomized argument that generates the entries of $H$ independently according to a uniform distribution in $\{0,1\}$, we can show that there exists an $H$ with $\Psi(H)=0$ for sufficiently large $n$ (see Appendix~\ref{App:3} for details). For such an $H$, we have from \eqref{BB} and \eqref{AA} that
$(1/n)I(M_1,M_2;Z^n) \leq 3/(n\epsilon)$ when the size of the realization of $S$ is $n\alpha_2$, and
$(1/n)I(M_1;Z^n) \leq 3/(n\epsilon)$ when the size of the realization of $S$ is $n\alpha_1$.

Letting $n \rightarrow \infty$ and $\epsilon \rightarrow 0$ (in that order) proves the achievability of the rate pair $(1-\alpha_1,\alpha_1-\alpha_2)$ and hence completes the proof of the theorem. \hfill $\square$

\section{Concluding Remarks} \label{sec:con}
In this paper we considered the problem of simultaneously communicating two messages, a high-security message and a low-security message, to a legitimate receiver, referred to as the security embedding problem. An information-theoretic formulation of the problem was presented. With appropriate coding architectures, it was shown that a significant portion of the information bits can receive additional security protections without sacrificing the overall rate of communication. Key to achieve efficient embedding was to use the low-security message as part of the transmitter randomness to protect the high-security message when the eavesdropper channel realization is strong. 

For the engineering communication scenarios with real channel input and additive white Gaussian noise, it was shown that the high-security message can be embedded into the low-security message at full rate without incurring any loss on the overall rate of communication for both scalar and independent parallel Gaussian channels (under an average individual per-subchannel power constraint). The scenarios with multiple transmit and receive antennas are considerably more complex and hence require further investigations.

Finally, note that even though in this paper we have only considered providing two levels of security protections to the information bits, most of the results extend to multiple-level security in the most straightforward fashion. In the limit when the security levels change continuously, the number of secure bits delivered to the legitimate receiver would depend on the realization of the eavesdropper channel even though such realizations are  unknown a priori at the transmitter.

\appendix

\section{Proof of Theorem~\ref{thm:DM2}} \label{App:1}
First note that when $X \rightarrow Y \rightarrow Z_2$ forms a Markov chain in that order, we have $I(U;Y) \geq I(U;Z_2)$ for any jointly distributed $(U,V,X)$ that satisfies the Markov chain $U \rightarrow V \rightarrow X \rightarrow (Y,Z_1,Z_2)$. 

To show that the sufficient condition \eqref{eq:DM1-2} is also necessary, let $(R_1,R_2)$ be an achievable rate pair. Following Fano's inequality \cite{CT-B91} and the asymptotic perfect secrecy constraints \eqref{eq:cons2-1} and \eqref{eq:cons2-2}, there exists a sequence of codes  (indexed by the block length $n$) of rate pair $(R_1,R_2)$ such that 
\begin{eqnarray}
H(M_1,M_2|Y^n) & \leq & n\epsilon_n/2\label{eq:fano}\\
I(M_1;Z_1^n) & \leq & n\epsilon_n/2\label{eq:ct2}\\
\mbox{and} \quad I(M_1,M_2;Z_2^n) & \leq & n\epsilon_n/2\label{eq:ct3}
\end{eqnarray}
where $\epsilon_n \rightarrow 0$ in the limit as $n \rightarrow \infty$.

Following \eqref{eq:fano} and \eqref{eq:ct2}, we have
\begin{eqnarray*}
n(R_1-\epsilon_n) & = &H(M_1)-n\epsilon_n\\
& \leq & H(M_1) -\left[I(M_1;Z_1^n)+H(M_1,M_2|Y^n)\right]\\
& = & H(M_1|Z_1^n)-H(M_1,M_2|Y^n) \\
& \leq & H(M_1,M_2|Z_1^n)-H(M_1,M_2|Y^n) \\
& = & I(M_1,M_2;Y^n)-I(M_1,M_2;Z_2^n). 
\end{eqnarray*}
Let $M := (M_1,M_2)$, $Y^{i-1} := (Y[1],\ldots,Y[i-1])$, $Z_{1,i+1}^n=(Z_1[i+1],...,Z_1[n])$ and $U[i] := (Y^{i-1},Z_{1,i+1}^n)$. We further have
\begin{eqnarray*}
n(R_1-\epsilon_n) & \leq & I(M;Y^n) - I(M;Z_1^n)\\
& = & \sum_{i=1}^n \left[I(M;Y[i]|Y^{i-1}) -
I(M;Z_1[i]|Z_{1,i+1}^n)\right]\\
& \stackrel{(a)}= & \sum_{i=1}^n \left[I(M;Y[i]|Y^{i-1},Z_{1,i+1}^n) -
I(M;Z_1[i]|Y^{i-1},Z_{1,i+1}^n)\right]\\
& = &\sum_{i=1}^n \left[I(M;Y[i]|U[i]) -
I(M;Z_1[i]|U[i])\right]\\
& = & n\left[I(M;Y[Q]|U[Q],Q) -
I(M;Z_{1}[Q]|U[Q],Q)\right]\\
& = & n\left[I(M,U[Q],Q;Y[Q]|U[Q],Q) -
I(M,U[Q],Q;Z_{1}[Q]|U[Q],Q)\right]\\
& = & n\left[I(V[Q];Y[Q]|U[Q],Q) -
I(V[Q];Z_{1}[Q]|U[Q],Q)\right]
\end{eqnarray*}
where (a) is due to the Csisz\'{a}r-K\"{o}rner sum equality \cite[Lemma~7]{CK-IT78}, $Q$ is a standard time-sharing variable \cite{CT-B91}, and $V[Q]:=(M,U[Q],Q)$.

Following \eqref{eq:fano} and \eqref{eq:ct3}, we have
\begin{eqnarray*}
n(R_1+R_2-\epsilon_n) & = & H(M)-n\epsilon_n\\
& \leq &H(M)-\left[H(M|Y^n)+I(M;Z_2^n)\right]\\
& = & I(M;Y^n)-I(M;Z_2^n)\\
& = & \sum_{i=1}^n \left[I(M;Y[i]|Y^{i-1}) - I(M;Z_2[i]|Z_{2,i+1}^n)\right]\\
& \stackrel{(b)}= & \sum_{i=1}^n \left[I(M;Y[i]|Y^{i-1},Z_{2,i+1}^n) -
I(M;Z_2[i]|Y^{i-1},Z_{2,i+1}^n)\right]\\
& = & \sum_{i=1}^n
\left[I(M,Y^{i-1},Z_{1,i+1}^n,Z_{2,i+1}^n;Y[i]) -
I(M,Y^{i-1},Z_{1,i+1}^n,Z_{2,i+1}^n;Z_2[i])\right]\\
&& \quad\quad -\sum_{i=1}^n \left[I(Y^{i-1},Z_{2,i+1}^n;Y[i]) -
I(Y^{i-1},Z_{2,i+1}^n;Z_2[i]) \right]\\
&& \quad\quad -\sum_{i=1}^n \left[I(Z_{1,i+1}^n;Y[i]|M, Y^{i-1},Z_{2,i+1}^n) -
I(Z_{1,i+1}^n;Z_2[i]|M,Y^{i-1},Z_{2,i+1}^n)\right]\\
& \stackrel{(c)}\leq & \sum_{i=1}^n
\left[I(M,Y^{i-1},Z_{1,i+1}^n,Z_{2,i+1}^n;Y[i]) -
I(M,Y^{i-1},Z_{1,i+1}^n,Z_{2,i+1}^n;Z_2[i])\right]\\
& \stackrel{(d)}= & \sum_{i=1}^n \left[I(M,Y^{i-1},Z_{1,i+1}^n;Y[i]) -
I(M,Y^{i-1},Z_{1,i+1}^n;Z_2[i])\right]\\
& = & \sum_{i=1}^n \left[I(M,U[i];Y[i]) - I(M,U[i];Z_2[i])\right]\\
& = & n\left[I(M,U[Q];Y[Q]|Q) -I(M,U[Q];Z_{2}[Q]|Q)\right]\\
& = & n\left[I(M,U[Q],Q;Y[Q]) -I(M,U[Q],Q;Z_{2}[Q])-\left(I(Y[Q];Q)-I(Z_2[Q];Q)\right)\right]\\
& = & n\left[I(V[Q];Y[Q]) -I(V[Q];Z_{2}[Q])-\left(I(Y[Q];Q)-I(Z_2[Q];Q)\right)\right]\\
& \stackrel{(e)}\leq & n\left[I(V[Q];Y[Q]) -I(V[Q];Z_{2}[Q])\right]
\end{eqnarray*}
where (b) follows from the Csisz\'{a}r-K\"{o}rner sum equality \cite[Lemma~7]{CK-IT78},
(c) is due to the Markov chain \eqref{eq:mk2}, (d) is due to the Markov chain \eqref{eq:mk}, and (e) follows again from the Markov chain \eqref{eq:mk2} and the fact that the channel is memoryless.

Finally, we complete the proof of the theorem by letting $U:=(U[Q],Q)$, $V:=V[Q]$, $X:=X[Q]$, $Y:=Y[Q]$, $Z_1:=Z_1[Q]$, $Z_2:=Z_2[Q]$ and $n \rightarrow \infty$.

\section{Proof of Theorem~\ref{thm:DM3}} \label{App:2}
As shown in Theorem~\ref{thm:DM2}, when we have the Markov chains \eqref{eq:mk} and \eqref{eq:mk2}, there exists a random triple $(U,V,X)$ satisfying the Markov chain $U \rightarrow V \rightarrow X \rightarrow (Y,Z_1,Z_2)$ and such that $R_1 \leq I(V;Y|U)-I(V;Z_1|U)$ and $R_1+R_2 \leq I(V;Y)-I(V;Z_2)$. In fact, the sum rate $R_1+R_2$ can be further bounded from above as
\begin{eqnarray*}
R_1+R_2 & \leq & I(V,X;Y)-I(V,X;Z_2)-\left[I(X;Y|V)-I(X;Z_2|V)\right]\\
& \stackrel{(a)}= & I(X;Y)-I(X;Z_2)-\left[I(X;Y|V)-I(X;Z_2|V)\right]\\
& \stackrel{(b)} \leq & I(X;Y)-I(X;Z_2)
\end{eqnarray*}
where (a) follows from the Markov chain $V\rightarrow X\rightarrow (Y,Z_2)$, and (b) follows from the Markov chain \eqref{eq:mk2} so $I(X;Y|V) \geq I(X;Z_2|V)$.

When we further have the Markov chain \eqref{eq:mk3}, $R_1$ can be further bounded from above as
\begin{eqnarray*}
R_1& \leq & I(U,V;Y)-I(U,V;Z_1|U)-\left[I(U;Y)-I(U;Z_1)\right]\\
& \stackrel{(c)}= & I(V;Y)-I(V;Z_1) - \left[I(U;Y)-I(U;Z_1)\right]\\
& \stackrel{(d)}\leq & I(V;Y)-I(V;Z_1)\\
& = & I(V,X;Y)-I(V,X;Z_1)-\left[I(X;Y|V)-I(X;Z_1|V)\right]\\
& \stackrel{(e)}\leq & I(V,X;Y)-I(V,X;Z_1)\\
& \stackrel{(f)}= & I(X;Y)-I(X;Z_1)
\end{eqnarray*}
where (c) follows from the Markov chain $U \rightarrow V \rightarrow (Y,Z_1)$, (d) and (e) follow from the Markov chain \eqref{eq:mk3} so $I(U;Y) \geq I(U;Z_1)$ and $I(X;Y|V)\geq I(X;Z_1|V)$, and (f) follows from the Markov chain $V \rightarrow X \rightarrow (Y,Z_1)$. This completes the proof of the theorem.

\section{Existence of an $H$ with $\Psi(H)=0$} \label{App:3}
To show that there exists a parity-check matrix $H$ such that $\Psi(H)=0$, it is
sufficient to show that $\mathbb{E}\Psi(H)< 1$ where $\mathbb{E}X$ denotes the
expectation of a random variable $X$.

Let 
\begin{equation*}
\Psi_0(H) := \left\{
\begin{array}{rl}
1, & \mbox{rank}(H)< n(1-\alpha_2-\epsilon)\\
0, & \text{otherwise}
\end{array}\right.
\end{equation*}
and 
\begin{equation*}
\Psi_i(H,\Gamma) := \left\{
\begin{array}{rl}
1, & D_i(\Gamma)<n(1-\alpha_i-\epsilon)-3/\epsilon\\
0, & \text{otherwise}
\end{array}\right.
\end{equation*}
for $i=1,2$. By the union bound, we have 
\begin{equation}
 \mathbb{E}\Psi(H)\leq
\mathbb{E}\Psi_0(H)+\sum_{i=1}^2\sum_{\substack{\Gamma\subseteq
\{1,\ldots,n\}
\\|\Gamma|=n\alpha_i}}\mathbb{E}\Psi_i(H,\Gamma). 
\label{eq:App3-1}
\end{equation}
By \cite[Lemma~6]{OW-BSTJ84},
\begin{equation}
\mathbb{E}\Psi_0(H) \leq \frac{n(1-\alpha_2-\epsilon)2^{-n(\alpha_2+\epsilon)
} } { 1-2^{-n(\alpha_2+\epsilon)}} < \frac{1}{2}
\label{eq:App3-2}
\end{equation}
for sufficiently large $n$.
By \cite[Lemma~5]{OW-BSTJ84}, for any $\Gamma \subseteq \{1,\ldots,n\}$ such that $|\Gamma|=n\alpha_i$
\begin{equation*}
\mathbb{E}\Psi_i(H,\Gamma) \leq 2^{-3n+
n(1-\alpha_i-\epsilon)} \leq 2^{-2n}.
\end{equation*}
Since the total number of different subsets of $\{1,\ldots,n\}$ is $2^n$, we have
\begin{equation}
\sum_{i=1}^2\sum_{\substack{\Gamma\subseteq
\{1,\ldots,n\}
\\|\Gamma|=n\alpha_i}}\mathbb{E}\Psi_i(H,\Gamma) \leq 2\cdot 2^n\cdot 2^{-2n} = 2^{-n+1} < \frac{1}{2}
\label{eq:App3-3}
\end{equation}
for $n>2$. Substituting \eqref{eq:App3-2} and \eqref{eq:App3-3} into \eqref{eq:App3-1} proves that 
$\mathbb{E}\Psi(H)< 1$ for sufficiently large $n$ and hence completes the proof.

\end{document}